\documentclass{article}  
\usepackage{spadre2008}
\usepackage{graphicx}
\frompage{000} \topage{000}                                              
\def\rightmark{pQCD description of elliptic flow and jet-quenching}
\def\leftmark{O. Fochler et al.}
 
\title{QCD plasma equilibration, elliptic flow and jet-quenching 
\\ -- phenomena of common origin} 
\authors{
{Oliver Fochler, Andrej El, Zhe Xu and 
Carsten Greiner %
}\\[2.812mm]
{\normalsize
\hspace*{-8pt}Institut f\"ur Theoretische Physik, Goethe-Universit\"at,\\ 
 Max-von-Laue-Str.~1, D-60438 Frankfurt am Main, Germany
}}
 
\abstract{Fast thermalization, a subsequent 
strong buildup of elliptic flow of QCD matter 
and jet-quenching as found at the
Relativistic Heavy Ion Collider (RHIC) are understood as the consequence
of perturbative QCD (pQCD) interactions within the 3+1 dimensional parton
cascade BAMPS. The main contributions stem from pQCD inspired bremsstrahlung.
Comparisons to Au+Au data of the flow parameter 
$v_2$ 
as a function of participation number as well as the gluonic contribution to 
the nuclear modification factor $R_{AA}$
for most central collisions are given. 
Also the shear viscosity to entropy ratio is dynamically
extracted, which lies in the range of
$0.08$ and $0.15$, depending on the chosen coupling
constant.
}
 
\begin{document}
 
\maketitle
\setcounter{page}{1}
The values of the elliptic flow parameter $v_2$ measured by the experiments at
the Relativistic Heavy Ion Collider (RHIC) \cite{rhicv2} are (nearly) as large as those obtained from
calculations employing ideal hydrodynamics.
This finding suggests that a fast local equilibration of quarks
and gluons occurs at a very short time scale $\le 1$ fm/c.
and that the locally thermalized state of
matter created, the quark gluon plasma (QGP), behaves as a nearly perfect
fluid exhibiting strong ``explosive'' collective motion. Quarks and gluons
should be rather strongly coupled,
pointing towards a small viscosity to entropy coefficient for the QGP.
Besides the strong collective flow, 
jet--quenching has been observed at
RHIC
\cite{quench} as the second striking new
discovery. 
So far, both phenomena could not be related by a common understanding
of the underlying microscopic processes:
Jet--quenching 
is widely thought to be described 
by investigating the potential energy loss on the partonic level 
in terms of elastic pQCD collisions
and pQCD radiative processes.
 On the other hand, the strong elliptic flow implies early thermalization 
and liquid like bulk
properties which can not be understood by
binary pQCD collisions being considerably too weak.
This has raised the speculation about nonperturbative interactions
as well as about super symmetric representations of Yang-Mills theories
using the AdS/CFT conjecture.

In contrast, in this talk,
it is demonstrated that perturbative QCD  can still explain a fast
thermalization of the initially nonthermal gluon system 
\cite{XG05,XG07,EXG08}, the large collective
effects of QGP created at RHIC \cite{XGS08}, 
the smallness of the shear viscosity 
to entropy ratio \cite{XG08,XGS08}
and also the jet-quenching of high momentum partons \cite{FXG08}
in a consistent manner by using a relativistic pQCD based
on-shell parton cascade Boltzmann approach of multiparton scatterings
(BAMPS) \cite{XG05,XG07}. 

BAMPS is a parton cascade, which solves the Boltzmann transport equation
and can be applied to study, on a semi-classical level, the dynamics of
gluon matter produced in heavy ion collisions at RHIC energies. 
The structure of BAMPS is based on
the stochastic interpretation of the transition rate \cite{XG05},
which ensures full detailed balance for multiple scatterings. BAMPS subdivides
space into small cell units where the operations for transitions are performed.
Gluon interactions included in BAMPS are elastic and screened 
Rutherford-like pQCD $gg\to gg$ 
scatterings as well as pQCD inspired bremsstrahlung
$gg\leftrightarrow ggg$ of Gunion-Bertsch type. 
The matrix elements are discussed in the literature \cite{XG05,XG07,FXG08}.
The suppression of the bremsstrahlung
due to the Landau-Pomeranchuk-Migdal (LPM) effect is taken into account
within the Bethe-Heitler regime employing a step function
in the infrared regime, allowing for independent gluon emissions.

In the present simulations,
the initial gluon distributions are taken in a Glauber geometry 
as an ensemble of minijets with
transverse momenta greater than $1.4$ GeV \cite{XG07}, produced via
semihard nucleon-nucleon collisions.
The later interactions of the gluons are
terminated when the local energy density drops below $1\ \rm{GeV/fm}^3$.
This value is assumed to be the critical value for the occurrence of 
hadronization, below which parton dynamics is not valid. Because
hadronization and then hadronic cascade are not yet included in BAMPS,
a gluon, which ceases to interact, propagates freely and can be
regarded as a free pion employing a picture of parton-hadron duality.
(Implementing a Cooper-Frye prescription for hadronization and employing
UrQMD for the hadronic cascade are in progress.)
The minijet initial conditions and the subsequent evolution using
the present prescription of BAMPS for two sets of the coupling
$\alpha_s=0.3$ and $0.6$ give nice agreements to the measured 
transverse energy per rapidity over all rapidities
\cite{XGS08}.

 \begin{figure}[htb]
\vspace*{-.0cm}
\centering
\includegraphics[angle=0,width=0.7\textwidth]{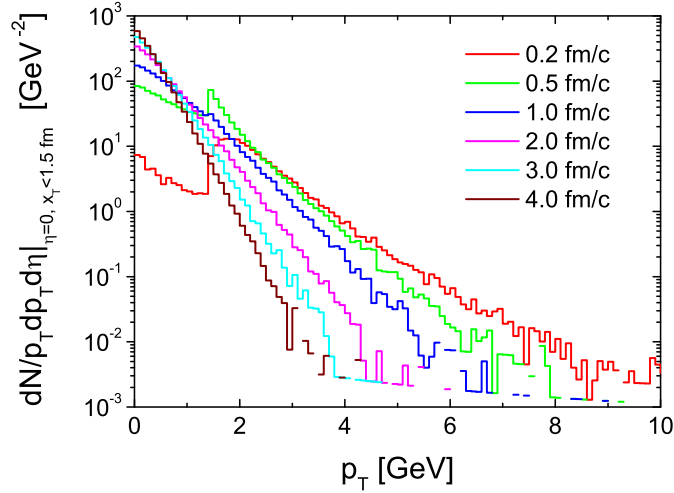}
\vspace*{-.0cm}
\caption[]{(Color online) Transverse momentum spectrum in the central region at
different times obtained from the BAMPS simulation
of a central Au+Au collision.}
\label{dndpt}
\end{figure}

As a first example  the fast thermalization of gluons is demonstrated 
in a local and 
central region which is taken as an expanding cylinder with a radius
of $1.5$ fm and within an interval of space time rapidity
$-0.2 < \eta < 0.2$. Figure \ref{dndpt} shows the varying transverse momentum
spectrum at various early times 
obtained from the BAMPS calculations for central Au+Au
collisions at $\sqrt{s}=200$ AGeV. The coupling constant is taken as
$\alpha_s=0.3$. 
In contrast of employing only binary pQCD collisions, the spectrum reaches an
exponential shape at $1$ fm/c and becomes increasingly steeper at late
times \cite{XG05,XG07}.
This is a clear indication for the achievement of local thermal
equilibrium and the onset of (quasi-) hydrodynamical collective expansion with
subsequent cooling by longitudinal work.

As a note aside, we remark that, in addition, potential color
instabilities \cite{instab} may play a role in isotropization of particle
degrees of freedom at the very initial stage where the matter is super dense.
However, more quantitative studies are needed to determine their significance
on the true thermal equilibration as suggested for the expanding quark gluon
matter at RHIC.

The inelastic pQCD based bremsstrahlung and its back reaction are
essential for the achievement of local thermal equilibrium at a short
time scale. The fast thermalization happens in a similar way if color
glass condensate is chosen as the initial conditions.
One of the important messages obtained there is that the hard gluons
thermalize at the same time as the soft ones due to the $ggg\to gg$
process, which is not included in the
so called standard ``Bottom Up'' scenario of thermalization \cite{EXG08}.

Kinetic equilibration relates to momentum deflection. Large momentum
deflections due to large-angle scatterings will speed up kinetic
equilibration enormously. Whereas the elastic pQCD scatterings favor
small-angle collisions, the collision and emmission angles in bremsstrahlung
processes are, for lower invariant, i.e. thermal energies,
rather isotropically distributed due to
the incorporation of the LPM cutoff \cite{XG05,XG07}. 
Hence, although the elastic cross section is still
considerably larger than the inelastic 
one, the given argument is the
intuitive reason why the bremsstrahlung processes are acting
more effectively
in the equilibration of the gluon matter than the elastic interactions.
Quantitatively it was shown in detail in \cite{XG07} that the
contributions of the different processes to momentum isotropization are
quantified by the so called transport rates
$$
R^{\rm tr}_i= \frac{\int \frac{d^3p}{(2\pi)^3} \frac{p_z^2}{E^2} C_i -
\langle \frac{p_z^2}{E^2} \rangle \int \frac{d^3p}{(2\pi)^3} C_i}{n\,
(\frac{1}{3}- \langle \frac{p_z^2}{E^2} \rangle)} \,,
$$
where $C_i[f]$
is the corresponding collision term describing various interactions,
$i=gg\to gg, gg\to ggg, ggg\to gg$, respectively.
The sum of them gives exactly the inverse of the time scale of momentum
isotropization, which also marks the time scale of overall thermalization.
As it turns out,
either by direct calculation or by the simulation, 
$R^{\rm tr}_{gg\to ggg}$ is a factor of $3-5$ larger than
$R^{\rm tr}_{gg\to gg}$ over a wide range in the coupling constant, 
which demonstrates the essential role of the
bremsstrahlung for thermal equilibration \cite{XG07}.
For a gluon gas, which is initially far away from equilibrium, one can
roughly estimate the time scale of thermalization $\tau_{\rm eq}$ by
taking the inverse of the sum of the transport collision rates close to
thermal equilibrium. At temperature $T=400$ MeV one has  
$\tau_{\rm eq}\approx 1/\sum R^{\rm tr}=0.32$ fm/c for $\alpha_s=0.3$.
We note that the above hinges on the assumption that the system is static.
Expanding systems are more complicated because particles flow, which
drives the system out of local equilibrium. Therefore, the momentum
degradation of flowing particles toward isotropy is slower than the
inverse of the total transport collision rate \cite{XG07}.

\begin{figure}[htb]
\vspace*{-.0cm}
\centering
\includegraphics[width=0.7\textwidth]{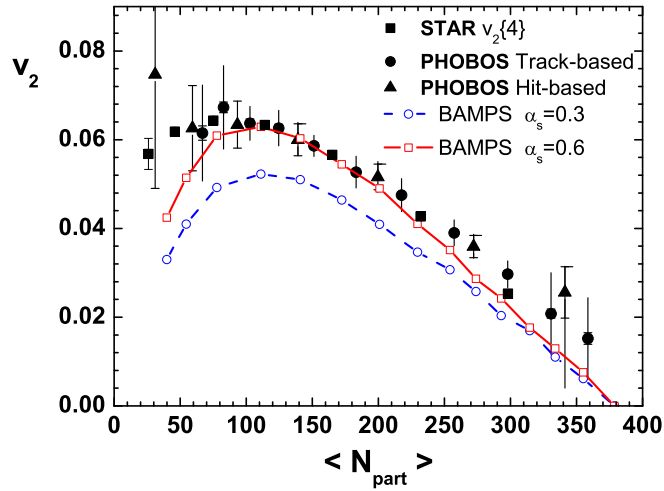}
\vspace*{-0cm}
\caption[]{(Color online) Elliptic flow $v_2(|y|<1)$ from BAMPS
using $\alpha_s=0.3$, and $0.6$, compared with the PHOBOS \cite{phobos}
and STAR \cite{star} data.}
\label{v2shv}
\end{figure} 

Further, employing the Navier-Stokes approximation the shear
viscosity $\eta$ is directly related to the transport rate \cite{XG08},
$$
\eta \cong \frac{1}{5} n \frac{\langle E(\frac{1}{3}-\frac{p_z^2}{E^2}) \rangle}
{\frac{1}{3}-\langle \frac{p_z^2}{E^2} \rangle} \frac{1}{\sum R^{\rm tr}+ 
\frac{3}{4} n \partial_t (\ln \lambda)}\,,
$$
where $\lambda$ denotes the gluon fugacity.
This expression allows to calculate the viscosity dynamically
and locally in a full and microscopical simulation.
On the other hand,
close to thermal
equilibrium the expression reduces to the more intuitive form 
$
\eta=\frac{4}{15} \, \frac{\epsilon}{\sum R^{\rm tr}}
$
and thus for the shear viscosity to entropy ratio 
$
\frac{\eta}{s}=
\left (5 \beta R^{\rm tr}_{gg \to gg}+ 
\frac{25}{3} \beta R^{\rm tr}_{gg \to ggg} \right )^{-1}
$.
Within the present description bremsstrahlung and its back reaction
lower the shear viscosity to entropy density ratio significantly by
a factor of $7$, compared with the ratio when only elastic collisions
are considered. For $\alpha_s=0.3$ one finds $\eta/s=0.13$
\cite{XG08,EXG08}. To match
the lower bound of $\eta/s=1/4\pi$ from the AdS/CFT
 conjecture $\alpha_s=0.6$ has to be employed.
Even for that case the cross sections are in the order of $1$ mb for a
temperature of $400$ MeV. Perturbative QCD interactions
can drive the gluon matter to a strongly coupled system with an
$\eta/s$ ratio as small as the lower bound from the AdS/CFT conjecture. 

The elliptic flow $v_2$ can be
directly calculated from microscopic simulations and compared to experimental
data, assuming parton-hadron duality.
$\alpha_s=0.3$ and $0.6$ are
used for comparisons \cite{XGS08}. Inspecting Figure \ref{v2shv},
except for the central centrality region the results with $\alpha_s=0.6$
agree perfectly with the experimental data, whereas the results with
$\alpha_s=0.3$ are roughly $20\%$ smaller. One can summarize that 
sufficient elliptic flow is being built up, yet
the details of the freeze-out, hadronization and
possible hadronic phase contributions might also give some minor dependence on
the overall strength in elliptic flow, which still have to be addressed.
In any case, for the QGP phase gluon bremsstrahlung dominates and
yields rapid thermalization, and, therefore, early pressure buildup,
and a small shear viscosity like a nearly ideal fluid.

\begin{figure}[htb]
\vspace*{-.0cm}
\centering
\includegraphics[width=0.7\textwidth]{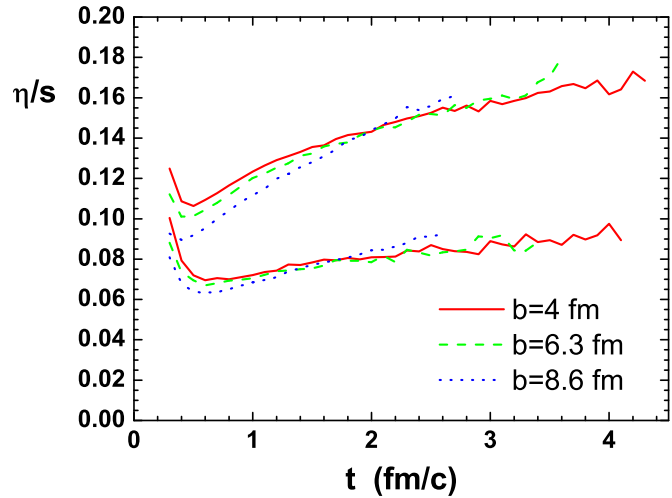}
\vspace*{-0cm}
\caption[]{(Color online) Shear viscosity to entropy density
ratio $\eta/s$ at the central region during the entire expansion. $\eta/s$
values are extracted from the simulations at impact parameter $b=4$,
$6.3$, and $8.6$ fm. The upper band shows the results with $\alpha_s=0.3$
and the lower band the results with $\alpha_s=0.6$. }
\label{shv}
\end{figure} 

Within these simulations
the ratio of the shear viscosity to the entropy density, $\eta/s$, 
can be locally extracted by means of the above formula.
This is shown in Figure \ref{shv}. 
As expected from the explicit expressions of the matrix elements,
the ratio
does not depend strongly on the gluon density or temperature, since
interaction rates and transport collision rates scale with the temperature.
Hence, $\eta/s$ depends practically only on $\alpha_s$. For $\alpha_s=0.6$,
at which the $v_2$ values match the experimental data, one has 
$\eta/s \approx 0.08$. However, $\eta/s$ may be higher, since inclusion of
hadronization and subsequent hadronic cascade 
can yield minor contributions to the final elliptic flow
values. Furthermore, different picture of initial conditions (eg. color
glass condensate) will also lead to different initially spatial eccentricity,
and, hence, will affect the final value of $v_2$ and the result on $\eta/s$.
These investigations are underway and will provide more constraints on
extracting $\eta/s$.

\begin{figure}[h]
\vspace*{-.0cm}
\centering
\includegraphics[width=0.7\textwidth]{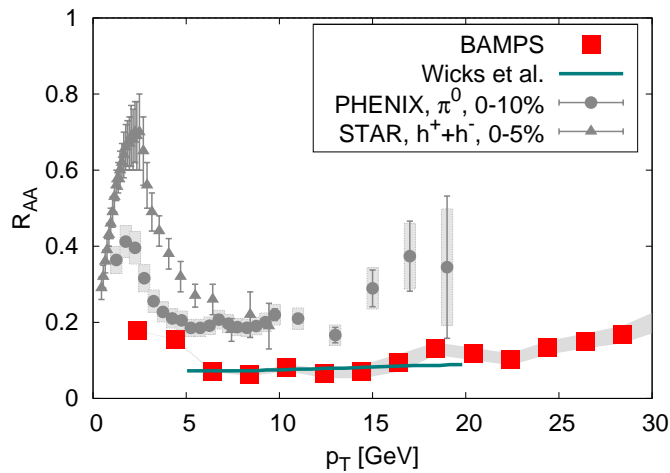}
\vspace*{-.0cm}
\caption[]{(Color online) Gluonic $R_{AA}$ at 
midrapidity ($y \,\epsilon\, [-0.5,0.5]$)
    as extracted from simulations for central Au+Au collisions at 200~AGeV.
    For direct comparison the result from
    Wicks et al. \cite{WHDG07} for the gluonic contribution to $R_{AA}$ and experimental results
    from PHENIX \cite{AA08} for $\pi^{0}$ and STAR \cite{JA03} for charged hadrons are shown.}
\label{fig:RAA}
\end{figure}  

The quenching of gluonic jets can now self-consistently be
addressed within the same full simulation of BAMPS \cite{FXG08}
including elastic and radiative collisions. 
Due to the steeply falling spectrum of initial mini--jets
a suitable weighting and reconstruction scheme has to be employed,
where a huge number of initial spectra are sampled which are then
selected for such events for further simulation that contain high--$p_{T}$
partons. The nuclear modification factor
$R_{AA}$ of the gluons is obtained directly by taking the ratio of the final $p_{T}$ spectra to the
initial mini-jet spectra.  Fig. \ref{fig:RAA} shows the result, 
exhibiting a clear
suppression of high--$p_{T}$ gluon jets at a roughly constant level
of $R_{AA}^{\mathrm{gluons}} \approx 0.1$, potentially slightly
rising towards high $p_{T}$.  The coupling constant is taken as
$\alpha_s=0.3$. 
The dominant contribution for the energy loss are the bremsstrahlung
contributions, although the picture changes for higher 
jet energies \cite{FXG08}.
The suppression is approximately a factor of two stronger than the
experimental pion data. This, however, was to be expected since at
present the simulation does not include quarks, which are expected
to lose less energy by a factor of $4/9$. Indeed, comparing with
state of the art results from Wicks et al. \cite{WHDG07} for
the gluonic contribution to $R_{AA}$ (seen as the line in Fig.
\ref{fig:RAA}), which in their approach together with the quark
contribution reproduces the experimental data, one finds a perfect
agreement. 

In summary, the pQCD based parton cascade BAMPS is used to
 calculate the time scale 
of thermalization, the elliptic flow $v_2$, extraction $\eta/s$, 
and the quenching of gluonic jets from simulations
of Au+Au collisions at RHIC energy $\sqrt{s}=200$ AGeV
within one common setup. This is a
committed and large scale undertaking. BAMPS includes
elastic $gg\to gg$ and inelastic bremsstrahlung and its back reaction
$gg \leftrightarrow ggg$.
The present
approach thus constitutes a first realistic 3+1 dim. microscopic
transport simulation that can describe the two phenomena
elliptic flow and jet-quenching
by the same
underlying processes. Further analyses on jet quenching,
on particle correlations, on quark degrees of
freedom including hadronization, on initial conditions and on the
exploration and use of dissipative hydrodynamics are underway to establish a more global picture of heavy ion collisions.

\section*{Acknowledgments}
We are grateful to the Center for the Scientific Computing (CSC) at Frankfurt for the computing resources. This work was supported by BMBF, DFG and GSI.

\vfill\eject
\end{document}